\documentstyle[epsfig]{aprim10}
\input{epsf}

\newif\ifAMStwofonts

\ifoldfss
  \ifCUPmtlplainloaded \else
    \NewTextAlphabet{textbfit} {cmbxti10} {}
    \NewTextAlphabet{textbfss} {cmssbx10} {}
    \NewMathAlphabet{mathbfit} {cmbxti10} {} 
    \NewMathAlphabet{mathbfss} {cmssbx10} {} 
  \fi
  \ifAMStwofonts
    \ifCUPmtlplainloaded \else
      \NewSymbolFont{upmath} {eurm10}
      \NewSymbolFont{AMSa} {msam10}
      \NewMathSymbol{\upi}     {0}{upmath}{19}
      \NewMathSymbol{\umu}     {0}{upmath}{16}
      \NewMathSymbol{\upartial}{0}{upmath}{40}
      \NewMathSymbol{\leqslant}{3}{AMSa}{36}
      \NewMathSymbol{\geqslant}{3}{AMSa}{3E}

    \fi
  \fi
\fi 

\ifnfssone
  \newmathalphabet{\mathit}
  \addtoversion{normal}{\mathit}{cmr}{m}{it}
  \addtoversion{bold}{\mathit}{cmr}{bx}{it}
  \newmathalphabet{\mathbfit} 
  \addtoversion{normal}{\mathbfit}{cmr}{bx}{it}
  \addtoversion{bold}{\mathbfit}{cmr}{bx}{it}
  \newmathalphabet{\mathbfss} 
  \addtoversion{normal}{\mathbfss}{cmss}{bx}{n}
  \addtoversion{bold}{\mathbfss}{cmss}{bx}{n}
  \ifAMStwofonts
    \ifCUPmtlplainloaded \else
      \UseAMStwoboldmath
      \makeatletter
      \new@mathgroup\upmath@group
      \define@mathgroup\mv@normal\upmath@group{eur}{m}{n}
      \define@mathgroup\mv@bold\upmath@group{eur}{b}{n}
      \edef\UPM{\hexnumber\upmath@group}
      \new@mathgroup\amsa@group
      \define@mathgroup\mv@normal\amsa@group{msa}{m}{n}
      \define@mathgroup\mv@bold\amsa@group{msa}{m}{n}
      \edef\AMSa{\hexnumber\amsa@group}
      \makeatother
      \mathchardef\upi="0\UPM19
      \mathchardef\umu="0\UPM16
      \mathchardef\upartial="0\UPM40
      \mathchardef\leqslant="3\AMSa36
      \mathchardef\geqslant="3\AMSa3E
    \fi
  \fi
\fi 

\ifnfsstwo
  \DeclareMathAlphabet{\mathbfit}{OT1}{cmr}{bx}{it}
  \SetMathAlphabet\mathbfit{bold}{OT1}{cmr}{bx}{it}
  \DeclareMathAlphabet{\mathbfss}{OT1}{cmss}{bx}{n}
  \SetMathAlphabet\mathbfss{bold}{OT1}{cmss}{bx}{n}
  \ifAMStwofonts
    \ifCUPmtlplainloaded \else
      \DeclareSymbolFont{UPM}{U}{eur}{m}{n}
      \SetSymbolFont{UPM}{bold}{U}{eur}{b}{n}
      \DeclareSymbolFont{AMSa}{U}{msa}{m}{n}
      \DeclareMathSymbol{\upi}{0}{UPM}{"19}
      \DeclareMathSymbol{\umu}{0}{UPM}{"16}
      \DeclareMathSymbol{\upartial}{0}{UPM}{"40}
      \DeclareMathSymbol{\leqslant}{3}{AMSa}{"36}
      \DeclareMathSymbol{\geqslant}{3}{AMSa}{"3E}
    \fi
  \fi
\fi 

\ifCUPmtlplainloaded \else
  \ifAMStwofonts \else 
    \def\upi{\pi}
    \def\umu{\mu}
    \def\upartial{\partial}
  \fi
\fi

\title[Rapid Binary Stellar Population Synthesis]{Rapid Binary Stellar Population Synthesis }

\author[Zhongmu Li]
       {Zhongmu Li\\
       Collage of Physics and electrical information, Dali University, Dali, 671003, China}
\date{}

\pagerange{\pageref{firstpage}--\pageref{lastpage}}
\pubyear{2008}

\begin{document}

\maketitle

\label{firstpage}

\begin{abstract}
Binary stars are common and it is necessary to model stellar
populations using binary stars. We introduce a method to model
binary-star stellar populations quickly. The method can also be used
to model single-star stellar populations. The method is called rapid
stellar population synthesis and it is based on a statistical
isochrone database of both single- and bianry-star stellar
populations. The database can be obtained easily from the CDS (or on
request to the authors) and can be used conveniently for most
stellar population studies and colour-magnitude fits. The main
feature of using such a technique for modeling stellar populations
is that it can save much computing time. Comparing to the method
which evolves binary-stars using stellar evolution codes directly,
the rapid stellar population synthesis method takes only one of
200,000 of the computing time. Although we can use the database to
model stellar populations quickly, the involved uncertainties in
spectra synthesis is small ($\sim$ 0.8\%). I also introduce the
method that uses the spectral energy distributions (SEDs) of single
stellar populations to model the SEDs of composite populations.
\end{abstract}


\section{Introduction}

Because galaxies contain a lot of stars and the evolution of stars
has been well understood. The technique of evolutionary population
synthesis has been an important tool for studying galaxies,
especially for studying their formation and evolution. Most works
used single-star simple stellar population (ssSSP) models. However,
there are many binaries in galaxies and binary interactions have
significant effects on stellar population studies. It is necessary
to model the populations of galaxies via both single and binary
stars.

Although there are many isochrone databases or star evolution tracks
that can be used for modeling ssSSPs, there is no suitable star
evolution tracks for modeling bianry-star stellar populations
(bsSSPs). Therefore, some previous works used a rapid binary
evolution code \cite{Hurley:2002} to calculate the evolution of
binaries, because the code can calculate the evolution of binaries
quickly. However, it usually needs a big star sample for modeling
bsSSPs, as we only know some distributions for the model inputs. In
this case, it usually takes long period for calculation of the
evolution of stars and then makes difficult to model stellar
populations via binaries widely. This is more significant when
modeling the evolution of galaxies. Aiming to solve the problem, an
isochrone database for both single-star and binary-star populations
was built and a rapid stellar population synthesis method was
presented \cite{Li:2008}.

The paper just aims to give a short introduction to the database of
bsSSPs and show the rapid spectral synthesis for stellar populations
and the rapid fitting for colour-magnitude diagrams (CMDs) of star
clusters.

\section{Rapid binary stellar population synthesis}\label{sec:sens}
\subsection{The basis of rapid binary stellar population synthesis}
The basis of rapid binary stellar population synthesis is the
isochrone database of bsSSPs \cite{Li:2008}. The main steps to build
the data base is as follows. First, a star sample of 2000, 000
binaries of populations was generated using a Monte Carlo method.
Second, the evolution of stars was calculated using rapid stellar
evolution code \cite{Hurley:2002} with default parameters of the
code. Third, the effective temperature ($T_{\rm eff}$, 2000-60000K)
versus surface gravity (log~$g$, -1.5-6) plane was divided into 1451
$\times$ 751 subgrids and the fraction of stars locating in each
subgrid was calculated and saved. Usually, for a population, there
are less then 20\,000 subgrids containing stars. Therefore, the
database gives the statistical distributions of stars of populations
in a log~$g$ versus $T_{\rm eff}$ plane. In order to make the
database more useful, it supplies the star distributions of
populations for both the initial mass functions of Salpeter
\cite{Salpeter:1955} and that of Chabrier \cite{Chabrier:2003}. The
input ranges for the metallicity and age of populations are
0.0001-0.03 and 0-15 Gyr, respectively.

Because the database supplies the distributions of stars of stellar
populations in the log~$g$ versus $T_{\rm eff}$ diagram, it is
unnecessary to calculate the evolution of stars when modeling
stellar populations. This can save a lot of computing time. In
addition, because the database supplies some statistical
distributions of stars on log~$g$ versus $T_{\rm eff}$ plane, rather
than the the positions of each star, when modeling the integrated
peculiarities of populations (e.g., calculating the spectral energy
distributions of populations, hereafter SEDs), the results of many
stars can be obtained via the same calculation. This can also speed
up the process of modeling populations.

\subsection{Modeling simple stellar populations}
The isochrone database can be directly used to calculate the spectra
or photometry of simple stellar populations, because the database
supplies very the statistical isochrones of simple stellar
populations. In fact, any spectral library and photometry library
can be used to translate the log~$g$ and $T_{\rm eff}$ of stars into
the spectra or photometries of populations easily.

\subsection{Modeling composite stellar populations}
Although early-type galaxies were thought to be close to simple
stellar populations and they were usually investigated by simple
stellar populations, it is better to study them via composite
stellar populations, because more and more observations showed
recent star formations in those galaxies. In order to get more
information about star formation histories of galaxies, it necessary
to build composite stellar population models. This can be easily
done via the isochrone database of SSPs or the SSP models presented
before. If one calculates the SEDs of composite populations via the
SEDs of ssSSPs, the computing time will be very short, but it will
take longer time to calculate SEDs from an isochrone database. In
addition, one can also use the SEDs of bsSSPs to calculate the ones
of composite binary populations. The SEDs of bsSSP can be found from
our previous work \cite{Li:2008}.

\subsection{Modeling colour-magnitude diagrams}
Colour-magnitude diagrams are powerful tools for studying the
stellar populations and other characteristics (e.g., binary
fraction, distance and colour excess) of star clusters. I introduce
the method of using the isochrone database of populations to model
the CMDs of star clusters. Because the isochrone database supplies
the log~$g$ and $T_{\rm eff}$ of stars of stellar populations, the
CMDs of stellar populations can be obtained via translating the
log~$g$ and $T_{\rm eff}$ values into the colours and absolute
magnitudes using a photometry library. Then one can fit the
observational CMDs using the CMDs of populations. The stellar
metallicities, ages, distances and colour excesses of star clusters
can be determined. The main feature of using the CMDs based on an
isochrone database to fit observational CMDs is its high fitting
speed and reliability. The high fitting speed results from the small
number of subgrids of populations. The high fitting reliability
results from the comparison of the fraction of stars in each
subgrid. In other words, when using this method to fit CMDs, the
shapes and star distributions are compared at the same time. The
results is actually more accurate compared to those determined only
by the shapes of CMDs. Because the uncertainties of different parts
of CMDs are different, it seems better to give different weights for
different parts. Furthermore, the shapes of CMDs are usually
affected by the fraction of binaries that are observed as single
stars. When explaining the CMDs of star clusters, the effects of the
fraction of undistinguishable binaries should be taken into account.

\subsection{Modeling the spectral and photometry evolution of galaxies}
In the studies about the formation and evolution of galaxies, the
modeling of spectral and photometry evolution is important. The
reason is that stars contribute mainly to the eyeable mass and the
light of galaxies. Here I introduce a little about using the
isochrone database of stellar populations to model the spectral and
photometry evolution of galaxies. Because there are too many stars
in a galaxy, using a statistical isochrone database instead of the
evolution tracks of various stars to model galaxies is convenient.
This will save much computing time. The main thought here is similar
to that of modeling the SEDs and photometry of composite stellar
populations, but the effects of dust and gas should be considered.
In addition, it seems better to use some spectral libraries
containing emission spectra, as there are many special activities
such as large fraction star formations when galaxies evolve.

\section{Conclusion and discussion}\label{sec:discuss}

I mainly present the applications of using a isochrone database of
binary-star stellar populations to model binary-star populations.
The application of the database and the SEDs of simple binary-star
populations to model the CMDs of star clusters and the spectral and
photometry evolution is also discussed. It is shown that the rapid
binary population synthesis can be widely used in astrophysics
studies, especially in the studies of galaxies.

\section*{Acknowledgment}
We are grateful to Profs. Zhanwen Han and Gang Zhao for useful
discussions and suggestions.

\label{lastpage}

\clearpage

\end{document}